\documentstyle[preprint,aps,prd]{revtex} 

\begin{document}

\preprint{DAMTP/96-93, gr-qc/9610075}

\title{Comment on ``Spacetime Information''}

\author{Adrian Kent\thanks{E-mail: apak@damtp.cambridge.ac.uk}}

\address{Department of Applied Mathematics and Theoretical Physics,\\
  University of Cambridge,\\ Silver Street, Cambridge CB3 9EW, U.K.}

\date{18th February, 1997}
\maketitle
\begin{abstract}
\noindent
A recent paper by Hartle\cite{Hartle:spacetimeinfo} 
proposes a definition of ``spacetime
information'' --- the information available about a quantum 
system's boundary conditions in the various sets of decohering 
histories it may display --- and investigates its properties. 
We note here that the analysis of Ref. \cite{Hartle:spacetimeinfo} 
contains errors which invalidate several of the conclusions.    
In particular, the proof that the proposed definition 
agrees with the standard definition for ordinary quantum mechanics is 
invalid, the evaluations of the spacetime information for time-neutral
generalized quantum theories and for generalized quantum theories with
non-unitary evolution are incorrect, and the argument 
that spacetime information is conserved on 
spacelike surfaces in these last theories is erroneous. 
We show however that the proposed definition does, in fact, agree with
the standard definition for ordinary quantum mechanics.  
Hartle's definition relies on choosing, case by case, a class of 
fine-grained consistent sets of histories.  We supply 
a possible general definition of a class of sets that includes all 
the sets considered in 
Ref. \cite{Hartle:spacetimeinfo} and that generalizes to other cases.  
\end{abstract}

\vspace{15pt} \pacs{PACS numbers: 03.65.Bz, 04.60.-m, 04.70.Dy, 98.80.Hw
 \\[10pt] Submitted to Phys. Rev. D}

In a characteristically intriguing recent
paper\cite{Hartle:spacetimeinfo}, Hartle investigates the problem of
characterising the information available about the boundary conditions
in generalized quantum theories in which the notion of a quantum state
on a spacelike surface is not necessarily defined.  
Hartle's definition has three main ingredients: the standard
definition for the missing information in a probability distribution,
the formulation of quantum theory in terms of consistent sets of 
histories, and Jaynes' maximum entropy construction\cite{jaynes}. 

Specifically, Hartle considers quantum theories defined by 
general decoherence functionals $D$, which encode the boundary
conditions and perhaps other information, and supposes that some
natural standard class ${\cal C}_{\rm stand}$ of consistent sets 
of histories has been fixed.  
He then defines the missing information in the boundary conditions 
to be 
\begin{equation}\label{one}
{\cal S}(D) = \mathop{\min}_{\{C_\alpha\}\in\ {\cal C}_{\rm
stand}}
\left[-\sum\nolimits_\alpha
p(\alpha)\log p(\alpha)\right]\ ,
\end{equation}
where $ p(\alpha)$ is the probability of the history
$ C_{\alpha}$.  
Given this definition, Hartle proceeds to define the missing
information in any consistent set of histories to be the maximum
missing information in any set of boundary conditions which 
reproduce the decoherence functional matrix elements applied to 
the set:  
\begin{equation}\label{two}
S(\{C_\alpha\}) = \mathop{\rm max}\limits_{\tilde D} \left[{\cal S}
(\tilde D)\right]_{\tilde D(\alpha^\prime, \alpha) = D(\alpha^\prime,
\alpha)} \, . 
\end{equation}
The missing information in a class ${\cal C}$ 
of consistent sets of histories is then defined to be 
\begin{equation}\label{three} 
S({\cal C}) = \mathop{\min}_{{\rm
decoherent}\atop \{C_\alpha\}\in {\cal C}} S(\{C_\alpha\})\ .
\end{equation}
Finally, the complete information available is defined to be the
missing information in the class of all consistent sets of histories: 
\begin{equation}\label{four} 
S_{\rm compl} = \mathop{\min}_{{\rm decoherent}\ \{C_\alpha\}}
S(\{C_\alpha\})\ .
\end{equation}

Obviously, these definitions are open to 
criticism.
Consistent sets of histories have some strange 
properties\cite{Dowker:Kent:properties,Dowker:Kent:approach,%
Kent:quasi,Kent:contra,Kent:implications} and it
does not seem at all clear that they are the most natural objects to 
choose from within a spectrum which ranges, at least, from the class
of all sets of quantum histories, through the sets of linearly 
positive histories\cite{Goldstein:Page}, to the ordered consistent sets of 
histories\cite{Kent:implications}. 
It does not seem clear, either, that the maximum value of the 
missing information 
{\em defined by} any possible set of boundary conditions is 
a particularly useful measure of the missing 
information {\em about} the boundary conditions. 

Hartle's proposals might thus perhaps best be seen as a 
pioneering attempt to investigate a question which is still in 
need of some elucidation. 
It would be interesting to understand more clearly what precisely 
it is that we are attempting to measure when we frame a definition 
of spacetime information, and (hence) what properties should or might 
be required of the measure.   
It may well be that there are several different natural measures, 
suited for different purposes.  Indeed, Isham and Linden have
recently proposed alternative definitions\cite{Isham:Linden} which apply 
within their algebraic generalisation of the consistent histories
formalism, while Gell-Mann and Hartle\cite{hartlepriv} 
and the author\cite{kent:info} have also investigated 
other possibilities. 

Still, Hartle's proposed measure of missing information is a 
relatively simple
quantity which gives at least some indication of how well the 
boundary conditions are constrained.  It has also been
investigated as a possible ingredient in a set selection 
mechanism\cite{CEPI:GMH,McElwaine:2}.
In the remainder of this Comment we accept 
definitions (\ref{one}--\ref{four}) and examine their properties.  

We need first to define the class ${\cal C}_{\rm stand}$.  
After discussing two possible general definitions,
neither of which seems to be completely satisfactory, in Section VI
of Ref. \cite{Hartle:spacetimeinfo}, Hartle resorts to
supplying definitions case by case.  We would here
like to suggest a new general definition that might be of use.  

To define this class, we need to
assume that the theory admits a natural class of completely 
fine-grained projective decompositions of the identity: that is,
decompositions of the identity into orthogonal one-dimensional 
projections.  Let us call these {\it finest decompositions}. 
The specific examples we have in mind here are sets of one-dimensional
projections applied at any single time in a non-relativistic
theory, such as standard quantum mechanics or the time-neutral
quantum mechanics considered by Gell-Mann and Hartle\cite{gmhthree}, 
or applied on any spacelike hypersurface in a theory with fixed 
background geometry, such as those modelled by Anderson\cite{anderson}.
We need also to assume that at least one member of this class defines
a consistent set of histories.

When these assumptions hold, as they do in the examples considered
below, we propose to take ${\cal C}_{\rm stand}$ to be 
the class of consistent sets of histories that are defined entirely by
finest decompositions and that cannot consistently be extended by 
any further finest decompositions.  
This definition of ${\cal C}_{\rm stand}$ includes all the 
consistent sets allowed by Hartle's definitions 
in the examples considered in Ref. \cite{Hartle:spacetimeinfo}.  
It also applies, for example, to the case of a background 
geometry with more than one compact non-chronal region, for which 
no definition seems so far to have been suggested. 
However, as we will see below, it also 
includes consistent sets disallowed by Hartle's
definitions in the case of a background geometry with a non-chronal
region. 

We turn now to Hartle's analysis of the properties of 
definitions (\ref{one}--\ref{four}).  It is helpful to begin with 
the discussion of time-neutral generalized quantum mechanics in 
Section V, since this causes difficulties elsewhere in the paper. 
Following the above scheme, Hartle defines 
\begin{equation}
{\cal S} \bigl(\tilde\rho^f, \tilde\rho^i\bigr) \equiv
\mathop{\min}_{{\rm fine-grained}\atop{\rm decoherent}
\ \{C_\alpha\}}
\left[-\sum\nolimits_\alpha p(\alpha)\log p(\alpha)\right]
\end{equation}
where the minimum is over the completely fine-grained decoherent
 sets $\{C_\alpha\}$ for which
\begin{equation}
D(\alpha^\prime, \alpha) = {\cal N}\ {\rm Tr}\left(\tilde\rho^f
C_{\alpha^\prime} \tilde\rho^i C^\dagger_\alpha\right) =
\delta_{\alpha^\prime\alpha} p (\alpha)\ .
\end{equation}
The completely fine-grained histories consist of sequences
of sets of one-dimensional projections at each and every time, 
but since repeating the same set of projections has no effect
on the entropy, only the distinct projective decompositions in any
given consistent set need be considered.  

Hartle argues that, since non-trivial consistent extensions 
increase the value of 
$-\sum\nolimits_\alpha p(\alpha)\log p(\alpha)$, the minimum 
is attained by a set of histories of the form 
\begin{equation}
\label{historypair}
C_\alpha = P^f_{\alpha_f} P^i_{\alpha_i} \, , 
\end{equation}
where $P^f_{\alpha_f}$ are projections onto a basis
$\{|\alpha_f\rangle\}$ diagonalizing $\tilde\rho^f$ and $P^i_{\alpha_i}$
are projections onto a basis $\{|\alpha_i\rangle\}$ diagonalizing
$\tilde\rho^i$.

This is wrong, for two reasons. 
First, the increasing 
entropy argument, given by Hartle in equations (5.8)--(5.10), 
actually implies that the 
minimum is attained 
by histories containing a single projective decomposition, 
repeated at all times, rather than two different decompositions.   
Second, although the diagonalization of $\tilde\rho^i$ and
$\tilde\rho^f$ by the relevant projections is a sufficient condition
for consistency, it is not necessary.  The minimum is attained by 
a set of histories of the form 
\begin{equation}
C_\alpha = P_{\alpha} \, , 
\end{equation}
and in general the projections $P_{\alpha}$ need not diagonalize either 
$\tilde\rho^i$ or $\tilde\rho^f$.
The consistency condition
\begin{equation}
\langle\alpha | \tilde\rho^i | \alpha' \rangle \langle \alpha' | 
\tilde\rho^f | \alpha \rangle =  {\rm Tr}\left(\tilde\rho^f P_{\alpha}
\tilde\rho^i P_{\alpha'} \right) = 0 \quad {\rm~if~} 
\alpha \neq \alpha'  
\end{equation}
requires only that for each pair of distinct $\alpha , \alpha'$
either $\langle \alpha |  \tilde\rho^i | \alpha' \rangle$ 
or $\langle \alpha |  \tilde\rho^f | \alpha' \rangle$
should vanish.  There are generally many bases with this property. 
For example, we could choose a basis $\{ | \alpha_i \rangle \}$ 
by taking $| \alpha_1 \rangle$ to be an eigenvector
of $\tilde\rho^i$, $| \alpha_2 \rangle$ to be an eigenvector of the matrix
obtained by restricting $\tilde\rho^f$ to the subspace orthogonal to 
$| \alpha_1 \rangle$, $| \alpha_3 \rangle$ to be an eigenvector of 
the matrix obtained by restricting $\tilde\rho^i$ to the subspace orthogonal
to $| \alpha_1 \rangle$ and $| \alpha_2 \rangle$, and so on. 

Most of the discussion in the remainder of Section V of
Ref. \cite{Hartle:spacetimeinfo} is therefore incorrect. 

There seems to be an additional problem later in this section. 
Hartle observes that 
\begin{equation}
\label{compl} 
S_{\rm compl}\geq {\cal S}(\rho^f, \rho^i) 
\end{equation}
by definition, and then goes on to suggest that the equation is an
equality when $\rho^i$ and $\rho^f$ are non-degenerate. 
The condition required for this argument to go through is that the 
decoherence matrix elements 
of a consistent set of histories of the form (\ref{historypair}) 
should determine the operators $\rho^i$ and $\rho^f$ up to rescalings, not 
that the operators should determine the set, so that the 
degeneracy of $\rho^i$ and $\rho^f$ seems to be irrelevant. 
It is, however, true that (\ref{compl}) is generically an equality. 
A sufficient condition for
$\rho^i$ and $\rho^f$ to be determined up to rescalings, 
which generically holds true, is that there exist 
bases $\{ | \alpha_i \rangle \} $ and 
$\{ | \beta_i \rangle \}$ diagonalizing $\rho^i$ and $\rho^f$ 
respectively, with the property that $ \langle \alpha_i | \beta_j
\rangle \neq 0$ for all $i,j$.  When these bases exist, the 
corresponding fine-grained projective decompositions define 
the desired consistent set.

It follows that, generically, it is true that complete information
is available about the initial and final conditions on any pair of
separated spacelike surfaces, as Hartle suggests in section V.D.
It does not seem immediately obvious, though, that 
complete information is {\it generically} not available on a 
single spacelike surface, as Hartle suggests, although it 
seems usually to be true that a single spacelike surface will not
suffice. 

The discussion of spacetime information in standard quantum mechanics,
which follows and relies on equation (5.13), is also invalid. 
To understand the point of this discussion, note that, while the 
aim of Ref. \cite{Hartle:spacetimeinfo} is to define a measure 
of spacetime information from first principles, the standard entropy 
functional, 
\begin{equation}
{\cal S} (\tilde\rho) = -{\rm Tr}\left(\tilde\rho\log\tilde\rho\right)
\, , 
\end{equation}
for a single density matrix is taken for granted in the text up to 
equation (5.16).  
Hartle notes this when introducing the entropy functional in equation
(2.6), and promises there a justification from a more general point
of view in Section V.  It is this justification which equations 
(5.13)--(5.16), and the accompanying text, are intended (inter
alia) to supply.  Hartle concludes, after equation (5.16), that the
standard entropy functional on single density matrices has 
been derived as the least missing information in fine-grained
decoherent sets of histories. 

In this discussion, Hartle considers the case 
when $\tilde\rho^f=I$ and $\tilde\rho^i\equiv
\tilde\rho$, and assumes that the fine-grained projective
decompositions which define consistent sets are precisely those 
which diagonalize the initial density matrix $\tilde\rho$.  

In fact, {\em any} projective decomposition at a single time defines 
a consistent set for these boundary conditions, and in particular 
all fine-grained
projective decompositions define consistent sets.  The minimization  
therefore needs to be carried out over all fine-grained projective
decompositions.  

In other words, the proof of another of the key claims of 
Ref. \cite{Hartle:spacetimeinfo}, namely the derivation of the 
usual information measure 
for single density matrices from Hartle's general definition, 
\begin{equation}
\label{measuresequal}
{\cal S}_{\rm usual} (\tilde\rho) = -{\rm Tr}\left(\tilde\rho\log\tilde\rho\right)
= {\cal S}_{\rm Hartle} (\tilde\rho) \, , 
\end{equation}
fails.  
It is, however, true that the two measures agree, 
as we now show.  (We generally write ${\cal S} (\tilde\rho )$ for both
these measures, since they are equal, but have added labels   
here and below for clarity.) 

According to the general definition of
Ref. \cite{Hartle:spacetimeinfo}, the missing information for a standard
quantum theory with the initial state $\tilde\rho$ is 
\begin{equation}
{\cal S}_{\rm Hartle} \bigl( \tilde\rho \bigr) \equiv
\mathop{\min}_{{\rm fine-grained}\atop{\rm decoherent}
\ \{C_\alpha\}}
\left[-\sum\nolimits_\alpha p(\alpha)\log p(\alpha)\right] \, , 
\end{equation}
where the minimum is over the completely fine-grained consistent 
 sets $\{C_\alpha\}$ for which
\begin{equation}
D(\alpha^\prime, \alpha) = {\cal N}\ {\rm Tr}\left(
C_{\alpha^\prime} \tilde\rho C^\dagger_\alpha\right) =
\delta_{\alpha^\prime\alpha} p (\alpha)\ .
\end{equation}
The increasing entropy argument means that it is sufficient to 
take the minimum over consistent sets defined by fine-grained 
projective decompositions $\{P_\alpha\}$.  
We therefore need to show that 
\begin{equation}
\mathop{\min}_{{\rm fine-grained} \{P_\alpha\}}
\left[-\sum\nolimits_\alpha p(\alpha)\log p(\alpha)\right] = 
- {\rm Tr}\left(\tilde\rho\log\tilde\rho\right) \, . 
\end{equation}
Consider any fine-grained projective decomposition $\{P_\alpha\}$,
write $\tilde\rho = (\tilde\rho)_{\alpha \beta}$ as a matrix
in the basis defined by $\{P_\alpha\}$, and define the 
density matrix $\tilde\rho'$ in the same basis by
\begin{equation} 
( \tilde\rho' )_{\alpha \beta } = 
\left\{ \begin{array}{ll}  (\tilde\rho )_{\alpha \beta } & 
\quad \mbox{for $\alpha = \beta$,} \\ 
0 & \quad \mbox{for $\alpha \neq \beta$.}
\end{array} \right.
\end{equation}
We have that $(\tilde\rho)_{\alpha \alpha} = (\tilde\rho')_{\alpha
  \alpha} = p ( \alpha )$, so that 
\begin{equation}
- {\rm Tr}\left(\tilde\rho\log\tilde\rho'\right) = 
- \sum\nolimits_\alpha p(\alpha)\log p(\alpha) \, . 
\end{equation}  
Now  
\begin{equation} 
- {\rm Tr}\left(\tilde\rho\log\tilde\rho'\right) \geq 
- {\rm Tr}\left(\tilde\rho\log\tilde\rho\right) 
\end{equation} 
holds\cite{Hartle:spacetimeinfo,bratteli:robinson}
for all density matrices $\tilde\rho$ and $\tilde\rho'$. 
On the other hand, when the decomposition $\{P_\alpha\}$ diagonalizes
$\tilde\rho$, we have that $\tilde\rho = \tilde\rho'$ and hence 
that 
\begin{equation} 
- {\rm Tr}\left(\tilde\rho\log\tilde\rho'\right) = 
- {\rm Tr}\left(\tilde\rho\log\tilde\rho\right) \, . 
\end{equation} 
Thus (\ref{measuresequal}) holds, as claimed. 

For completeness we note also that, in the discussion of standard
quantum mechanics in Section II of Ref. \cite{Hartle:spacetimeinfo}, 
equations (2.21) and (2.22) 
are unnecessary to establish (2.23), which holds by definition. 

We turn now to the discussion of Section VII of 
Ref. \cite{Hartle:spacetimeinfo}, which considers 
Anderson's proposal\cite{anderson} for modelling generalized quantum theories
in background spacetimes with a non-chronal region  
by a decoherence functional which incorporates non-unitary 
evolution: 
\begin{equation}
D\left(\beta^\prime, \alpha^\prime; \beta, \alpha\right) =
Tr\left[X^{-1} C_{\beta^\prime} X C_{\alpha^\prime} \rho
C^\dagger_\alpha X^\dagger C_\beta (X^\dagger)^{-1}\right] \, , 
\end{equation}
where the non-unitary operator $X$ is assumed invertible. 

We first comment briefly on the definition of the 
class ${\cal C}_{\rm stand}$.  
Hartle proposes that in this case ${\cal C}_{\rm stand}$
should consist of all consistent sets that are defined by 
fine-grained projective decompositions and that include at least
one decomposition both before and after the non-chronal region. 
The increasing entropy argument means that, given this definition,
it suffices to consider sets of histories of the form 
\begin{equation}
\label{ncpairs} 
C_\alpha = P^i_\alpha = |\alpha\rangle\langle\alpha|\ ;
\quad C_\beta =
P^f_\beta = |\beta\rangle\langle\beta| \, , 
\end{equation}
where $\{|\alpha\rangle\}$ and 
$\{|\beta\rangle\}$ are bases. 

The general definition of ${\cal C}_{\rm stand}$ proposed 
above includes all the histories of the form (\ref{ncpairs}), 
where the projections are repeated on a maximal collection of
spacelike surfaces before and after the non-chronal region. 
However, it also allows fine-grained sets of histories 
all of whose projections are before the 
non-chronal region, provided that these sets have 
no extension by fine-grained projective decompositions 
after the region, and vice versa. 
It might perhaps be thought desirable to exclude such sets.
On the other hand, the definition has the virtue of extending 
to the case of a spacetime with several non-chronal regions, 
whereas it may not generally be possible to find {\it any} 
consistent set which includes fine-grained projective decompositions
in all the separate chronal regions. 

We now consider Hartle's analysis, using the definition of 
${\cal C}_{\rm stand}$ proposed in Ref. \cite{Hartle:spacetimeinfo}. 
To the extent that Hartle's discussion relies on the arguments of 
Section V, it is invalid. 
For instance, it is not always true that the 
histories of the form (\ref{ncpairs}), 
in which the bases diagonalize $\rho$ and $X X^{\dagger}$
respectively, are adequate for the 
purpose of computing $S_X (\rho )$.
To give a simple example, if 
$\rho = | \alpha_1 \rangle \langle \alpha_1 |$ is pure,
and we take $| \beta_1 \rangle$ to 
be $a X | \alpha_1 \rangle$, where $a$ is a normalising factor, 
then any orthonormal bases $\{ | \alpha_i \rangle \}$ and
$\{ | \beta_i \rangle \}$ that include
$| \alpha_1 \rangle$ and $| \beta_1 \rangle$ 
define a consistent set of the form (\ref{ncpairs}). 
Any set defined in this way has history probabilities with zero entropy, 
since all the histories except one have zero probability. 
On the other hand, for general $X$, all sets defined by bases 
diagonalizing $\rho$ and $X X^{\dagger}$ have 
history probabilities with positive entropy. 

The discussion of spacetime information on single
surfaces also contains errors.  Equation (7.15) in 
Ref. \cite{Hartle:spacetimeinfo} is 
incorrect, and should be replaced by 
\begin{eqnarray}
\tilde D(\alpha^\prime, \alpha) &=& {\rm Tr}\left[X^{-1} X P_{\alpha^\prime}
\tilde\rho P_\alpha X^\dagger (X^{-1})^\dagger\right]\nonumber \\
&=& \left\langle\alpha^\prime |\tilde\rho|\alpha\right\rangle 
\left\langle\alpha | \alpha^\prime \right\rangle =
D(\alpha^\prime, \alpha) =
\left\langle\alpha^\prime|\rho|\alpha\right\rangle\ 
\left\langle\alpha | \alpha^\prime \right\rangle \, .
\end{eqnarray}
Clearly, it is not true that, as Hartle suggests, $\tilde\rho=\rho$ is 
the unique density matrix which reproduces the decoherence functional.
This means that Hartle's evaluation of the information available on
spacelike surfaces before the non-chronal region is invalid, so that
his argument for the equality (7.13) fails.
It follows from Hartle's equations (7.17) and (7.18), however,
that the equality (7.13) is, at least, generically true, assuming
(as Hartle does) that the operator $X$ is known {\em a priori}. 

The argument for another key claim of
Ref. \cite{Hartle:spacetimeinfo}, 
that the information is conserved on spacelike surfaces on either side
of the non-chronal region, likewise fails, and 
it does not seem obvious that this claim is generally
true.

\section*{Acknowledgments}

I would like to thank Chris Isham, Noah Linden and 
Jim McElwaine for helpful discussions and Jim Hartle for a critical
reading of the manuscript and useful comments.  This work was 
supported by a Royal Society University Research Fellowship.

\end{document}